\documentclass[conference]{IEEEtran}
\IEEEoverridecommandlockouts 
\usepackage[utf8]{inputenc}
\usepackage[T1]{fontenc}
\usepackage{tipa}

\usepackage{cite}

\usepackage{graphicx}
\usepackage{float}
\usepackage{placeins}
\usepackage{url}
\usepackage{acronym}
\usepackage{hyperref} 

\usepackage{tikz}

\newcommand\copyrighttext{%
  \footnotesize \textcopyright 2024 or 2025 IEEE. Personal use of this material is permitted.
  Permission from IEEE must be obtained for all other uses, in any current or future
  media, including reprinting/republishing this material for advertising or promotional
  purposes, creating new collective works, for resale or redistribution to servers or
  lists, or reuse of any copyrighted component of this work in other works.}
\newcommand\copyrightnotice{%
\begin{tikzpicture}[remember picture,overlay]
\node[anchor=south,yshift=10pt] at (current page.south) {\fbox{\parbox{\dimexpr\textwidth-\fboxsep-\fboxrule\relax}{\copyrighttext}}};
\end{tikzpicture}%
}

\begin{document}
\acrodef{ICT}{information and communication technologies}
\acrodef{IMF}{International Monetary Fund}
\acrodef{EEA}{European Economic Area}
\acrodef{EGDI}{E-Government Development Index}
\acrodef{OECD}{Organization for Economic Co-operation and Development}
\acrodef{MFA}{multi-factor authentication}

%
\title{Understanding Trust in Authentication Methods for Icelandic Digital Public Services}

\author{
\IEEEauthorblockN{Brynjólfur Stefánsson\IEEEauthorrefmark{1}, Ásta Guðrún Helgadóttir\IEEEauthorrefmark{1}, Martin Nizon-Deladoeuille\IEEEauthorrefmark{1}\IEEEauthorrefmark{2}, Helmut Neukirchen\IEEEauthorrefmark{1},  Thomas Welsh\IEEEauthorrefmark{1} } 
\IEEEauthorblockA{\IEEEauthorrefmark{1}University of Iceland, Reykjavík, Iceland / \IEEEauthorrefmark{2}INSA Lyon, France\\
Email: brs69@hi.is, astagh@hi.is, martin.nizon-deladoeuille@insa-lyon.fr, helmut@hi.is, tomwelsh@hi.is }
}

\maketitle
\copyrightnotice

\begin{abstract}



Digital public services have revolutionised citizen and private sector interactions with governments. Certain communities are strongly dependent on such digital services for ensuring the availability of public services due to geographical isolation or the presence of adverse geophysical and weather phenomena. However, strong and effective security is key to maintaining the integrity of public records and services yet also for ensuring trust in them. Trust is essential for user uptake, particularly given a global increase in data-protection concerns and a turbulent geopolitical security environment. In this paper, we examine the case of public trust in various forms of authentication for electronic identification in Iceland, which has high availability requirements for digital public services due to its unique and dynamic geophysical characteristics. Additionally, Iceland has historically low levels of institutional trust which may conflict with the requirement for an increased need for digital public services. Through surveying the Icelandic general public, we find that there is a high-level of trust in digital identification services across all demographics. We conclude with a discussion and future research challenges towards improving the effectiveness of authentication considering the diverse groups within Icelandic society, such as the rapidly increasing population of migrants and the large and dynamic population of tourists.
\end{abstract}


\IEEEpeerreviewmaketitle

\section{Introduction}
The public's trust in digital services is a vital aspect of today’s society as we move further and further into the digital age. The needs of the public tend to correlate with the advancement in digital services and the development of secure authentication methods alongside them.
The Nordic region is regarded as successful in generating high levels of public trust due to fair and well-functioning societal institutions, characteristics of a welfare state and perceived absence of corruption \cite{andreasson2017trust}. In 
Iceland, however, trust in the government, its services as well as trust in the
banks has been negatively affected by the financial crisis of
2008 \cite{johnsen2018public}. This low institutional trust contrasts with the high degree of interpersonal trust within Icelandic society, which is also common in similar Nordic countries \cite{domanski2021relation}.

The assistance of the \ac{IMF}
through a systematic program to recover after the financial crisis has been an important aspect in the process of trust restoration in Icelandic society \cite{baldursson2018restoring}. However, post-crisis recovery processes of trust in Icelandic institutions were prolonged and arduous \cite{johnsen2018public}. It is important to emphasise that the general public in Iceland is characterised by constant change. Since 2001, Iceland is part of the Schengen Area, and with opening the \ac{EEA}'s labour market in 2006 to the newly joined EU member states, the numbers of foreigners residing in Iceland significantly increased \cite{palsson201414} and the numbers of foreign nationals (both immigrants and tourists) in Iceland continue to grow. In January 2024, Iceland had a population of 383,726 residents, 16.6\% of whom were migrants \cite{staticeStatisticsIceland}, in addition to 2.2 million tourists who visited Iceland in 2023~\cite{ferdamalastofaNumbersForeign}. This overall population, therefore, consists of diverse sub-groups, each with varying values and cultural attitudes to institutional trust and trust in digital government services. Not supporting the values of distinct users has been identified as a challenge in the software engineering community as it may reduce the effectiveness of the software and even erode the rights of the users \cite{whittle2019your}. These constant shifts in society might impact user uptake and, therefore, the effectiveness of digital services which can raise a question of the need for increased societal security. 

\begin{figure*}
    \centering
    \includegraphics[scale=0.75]{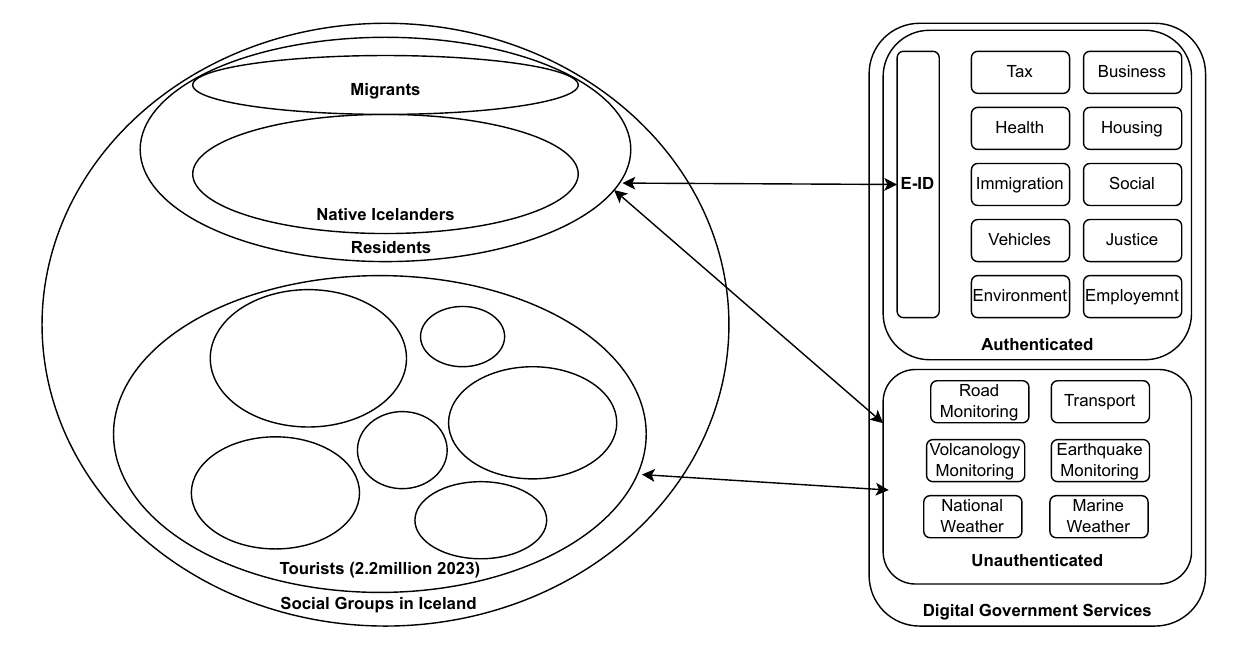}
    \caption{Diverse social groups and their need to access varying digital public services in Iceland.}
    \label{fig:ISgroups}
\end{figure*}

Digital government services are particularly important for societal resilience in Iceland which depends upon them to support communities given sparse population density and the presence of adverse geophysical events such as volcanic eruptions and extreme weather events. For example, road closures are dynamic in the face of rapidly changing weather and geology and are advertised online to both tourists and locals alike. As such, the need for enhancement of cybersecurity across all sectors to support resilience of critical infrastructures in Iceland has been recognised as a key priority \cite{govIS2022}. Figure~\ref{fig:ISgroups} illustrates our view on the diversity in social groups within Iceland who need to access various digital public services. To remain effective, authentication solutions must balance the needs of the diverse populations and the security of the nation.

This paper considers the need to understand the level of trust in authentication services for digital public services, and whether this trust is distinct from the low-level of institutional trust. Additionally, we consider how trust changes amongst population demographics as well as the willingness of the population to further improve upon those authentication methods, as the increasing need for cybersecurity and the evolution of authentication methods is on going. We achieve this through answering the following research questions:

\begin{description}
    \item [\textbf{RQ1}]How does trust in authentication for digital government services vary across demographics in the Icelandic population? 
    \item [\textbf{RQ2}]Which authentication methods have the highest level of trust? 
    \item [\textbf{RQ3}]Would the Icelandic general public be willing to add additional security layers into their authentication for digital public services?  
\end{description}

We take a quantitative approach to answering these research questions through means of an online survey directed at the Icelandic population.

The rest of this paper is structured as follows: Section~\ref{sec:Background} provides a background on trust in digital government services, Section~\ref{sec:Results and Analysis} presents the methodology and results, Section~\ref{sec:Discussion} discusses the results. Section~\ref{sec:Threats to Validity} covers possible limitations of the study. Section~\ref{sec:futureresearch} presents future research opportunities. Finally, Section~\ref{sec:Conclusion} concludes the paper with a summary and outlook.

\section{Background}\label{sec:Background}

\noindent\textbf{Digital government services} (or \emph{e-government}) pertain to the deployment of \ac{ICT} to facilitate service delivery both within governmental structures and between the government and its citizenry \cite{arias2018digital}. Over time, there has been a marked increase in the demand for seamless access to a plethora of services that traditionally necessitated in-person visits to pertinent establishments. This evolution underscores a societal shift towards prioritising convenience and efficiency in public service accessibility. 

Trust in digital government services can impact a society's institutional trust, inclusion and well-being \cite{ arias2018digital, janssen2021trustworthiness}. It can achieve this through increasing transparency which can support a reduction in corruption \cite{twizeyimana2019public}. The \ac{EGDI} measures the implementation of digital government services worldwide, illustrating that countries in Asia and Europe have strong success in its implementation~\cite{yuliantini2023impact}. However, challenges remain which centre around the variation in digital proficiency, or the \emph{digital divide}~\cite{sala2022gray}. Therefore, ensuring that digital government services are inclusive and accessible towards all members of the population is essential to enabling and sustaining institutional trust. However, in this context, it is necessary to distinguish between \emph{interpersonal} trust, i.e.\ trust between humans, and \emph{institutional trust}, i.e\ trusting private institutions or public organisations.

\noindent\textbf{Defining Trust.} The concept of trust can be understood in various ways. 
Rosseau et al. (1998, as cited in Moreno et al. ~\cite{moreno2014institutional}) defined trust as ``a psychological state comprising the intention to accept vulnerability based upon positive expectations of behaviour of another'' (p.~7). Rosseau’s understanding of trust emphasises an acceptance of putting oneself in a vulnerable position based on belief that the outcomes of others will be positive. In the context of trust in digital authentication methods this would mean that trusting governmental institutions guarantee security and that a person’s personal information is sheltered from any cyberthreats. Thus, as we focus on trust that locals in Iceland have towards different online authentication methods we are specifically looking at the ``raw'' characteristics of the concept of trust such as the ones suggested by Merriam-Webster dictionary \cite{mwtrust} ``to rely on the truthfulness or accuracy of, to place confidence in''. However we also consider that the element of vulnerability acceptance from Rosseau’s definition is rather significant, as individuals using authentication methods are at least to some extent familiar with cyberthreats. Taking these characteristics of trust, we analyse how the general public in Iceland perceives digital authentication methods and whether they feel that they can place their trust in them. 

\noindent\textbf{Trust Dynamics in Icelandic Society.} A recent study conducted by the \ac{OECD} in 2024 \cite{oecdpub} reveals that Icelanders exhibit one of the highest percentages of interpersonal trust at 82\%, and a substantial trust in the police at 73\%. Conversely, trust in the national government has experienced a decline to 36\% from its 50\% standing in 2021, placing it below the \ac{OECD} average of 39\%. Since the financial crisis of 2008, public trust in governmental institutions has been at historically low levels. Notably, prior to the parliamentary elections in 2021, there appeared to be a resurgence in governmental trust, which has since waned over the subsequent three years.
The Nordic region is acclaimed for fostering high levels of trust, attributed to equitable and efficient societal institutions, the hallmarks of welfare states, and a perceived absence of corruption \cite{andreasson2017trust}. Consequently, Icelandic society (as a Nordic nation) is emblematic of a high-trust society. However trust in governmental bodies, their services, and banking institutions was notably compromised following the financial crisis of 2008. The intervention by the \ac{IMF}, has been instrumental in the endeavour to restore trust of Icelandic society \cite{baldursson2018restoring}.

\noindent\textbf{Public Trust in Digital Services and Authentication.} The Icelandic population demonstrates elevated levels of trust among its populace towards one another, and even though the trust to governmental institutions is low the trust is higher towards law enforcement agencies, educational institutions, and administrative services~\cite{oecdpub}. The public's trust in digital services has become an essential facet of contemporary society, especially as the world advances further into the digital era. The evolving needs of the public are intrinsically linked to advancements in these services, necessitating the concurrent development of secure authentication methodologies. This is, in part, evident by the varied methods of authentication available for these services. Central to Icelandic society is the \emph{kennitala} or national identifier for personal or business use~\cite{watson2010short} which also functions as a means of authentication (e.g.\ given verbally) despite being highlighted for its transparency~\cite{watson2013unusually} which is unusual for this purpose. The rise of digital services (both public and private) have necessitated the use of more secure methods such as the \emph{rafræn skilríki} or electronic ID; where \ac{MFA} based on the kennitala can be deployed using a PIN as first factor and as second factor a smart card, a mobile app, or a Java applet on a SIM card. Across much of Europe, \ac{MFA} is becoming common, with Estonia being a notable example of success~\cite{lips2020eidas}. In Iceland, any private or public service can make use of this electronic ID for strong authentication. Given that some services have additional authentication requirements, such as an authentication app or bio-metrics, this can result in the several steps and multiple methods to authenticate which can both confuse and deter a user.  

Given the range of authentication methods available and the conflict between interpersonal trust and institutional trust in Iceland, we consider it necessary to evaluate societal trust in digital government services authentication methods to further understand their effectiveness in increasing institutional trust and the corresponding impact on their usability.

\section{Results and Analysis}\label{sec:Results and Analysis}
This study utilised a quantitative research methodology to research trust in authentication methods in Iceland and users' willingness to improve these solutions to improve security (RQ1-3). The authentication methods considered were: an authentication mobile app (Auth app), biometric, a digital certificate, in-person (i.e.\ visiting a premises), \ac{MFA} (i.e.\ the Icelandic electronic ID provided through a SIM card), a password manager, using a 3rd party social login (e.g.\ Google, Facebook) and a username/password combination.

The primary data was collected through an online survey  survey, administered via Microsoft Forms, distributed via the University of Iceland's social media platforms such as Facebook. The survey, designed collaboratively by the authors, employed scaled answer options to capture respondents attitudes towards the subject. Data collection occurred between June~1, 2024, and July~3, 2024, yielding a total of 379 responses. To determine the optimal sample size for this study, Cochran’s formula~\cite{ahmed2024} was used, given a 95\% confidence level and a 5\% margin of error. With an estimated Facebook-using population of 348,000 in Iceland \cite{napoleoncatFacebook2022}, a target sample size of 384 was calculated \cite{qualtricsSampleSize2023}. The survey’s 379 responses closely meet this ideal, making it sufficient for reliable analysis focused on adults in Iceland’s 2024 projected population of 383,726~\cite{staticeStatisticsIceland}.

The collected data was subsequently processed and analysed using the built-in features of Microsoft Forms, RStudio and Microsoft Excel. In the graphs shown, a scale from 1 to 5 is given where 1 represents no trust towards a certain method and 5 the most trust.

The data obtained from our online survey reveals that in Iceland, where a large part of the general public is already characterised by high levels of trust, individuals exhibit substantial trust towards various online authentication methods. The analysis of the data across different age groups, genders, and employment statuses indicates that the variations in trust are statistically insignificant. In the following subsections, we break down the results by demographic.

\subsection{Trust by age group}
\begin{figure}[ht]
    \centering
    \includegraphics[scale=0.5]{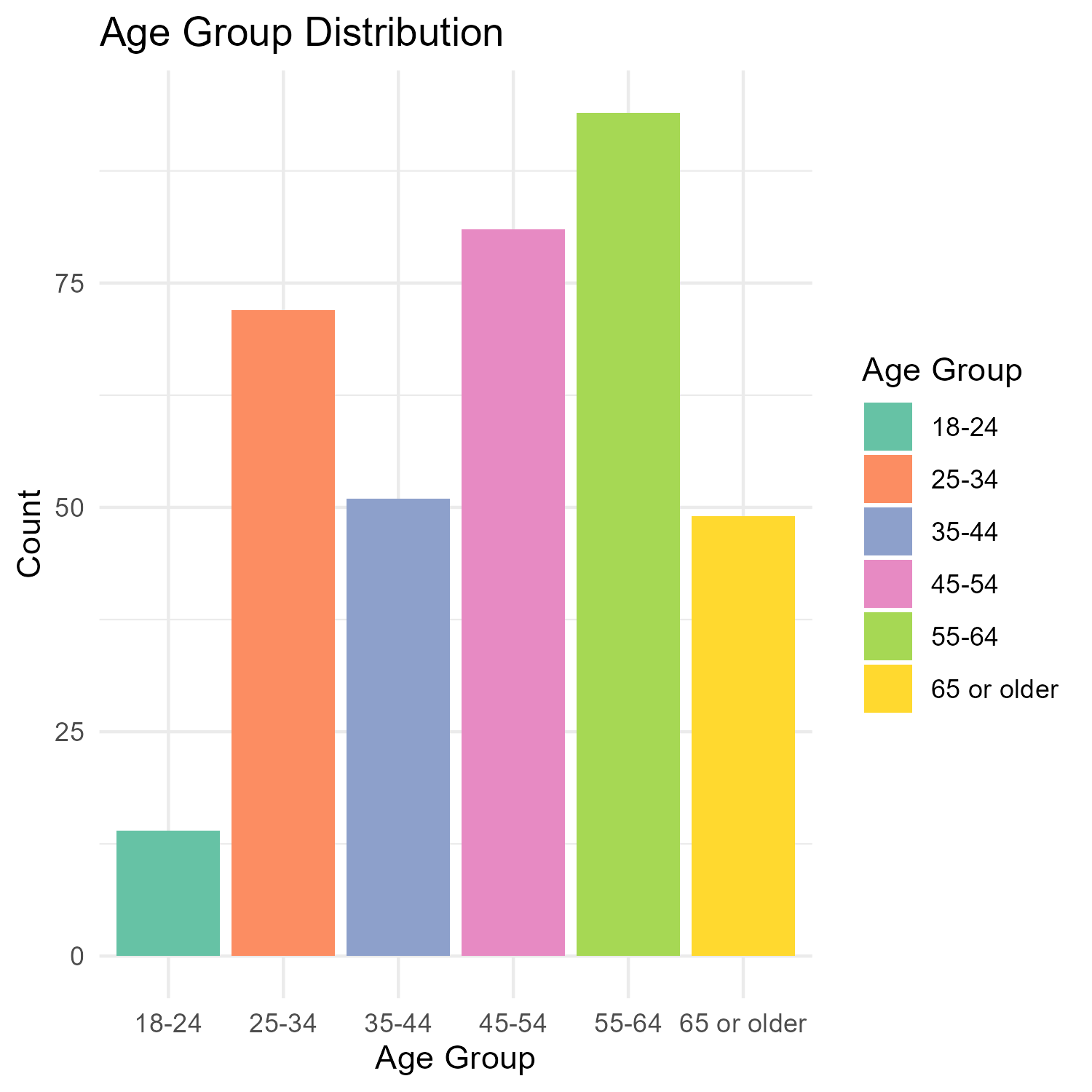}
    \caption{Age Group Distribution}
    \label{fig:ageDist}
\end{figure}

\noindent The distribution of different age groups was rather balanced (see Fig.~\ref{fig:ageDist}). However, the participation from people aged 24 and younger was significantly lower than anticipated whereas the participation of the age groups of 45-55 and 55-64 were greater than expected considering the age distribution of Facebook users in Iceland {\cite{napoleoncatFacebook2022}.  Figure~\ref{fig:trustByAge} presents the results of understanding trust across different age demographics. Younger age groups (\mbox{18-24}, \mbox{25-34}) tend to have more consistency in trust levels, particularly for biometrics, social login, and username/password methods. The older participants (\mbox{55-64}, 65 or older) show more variability  in their trust levels, with generally lower median trust levels across various methods. 

\begin{figure}[ht]
    \centering
    \includegraphics[scale=0.4]{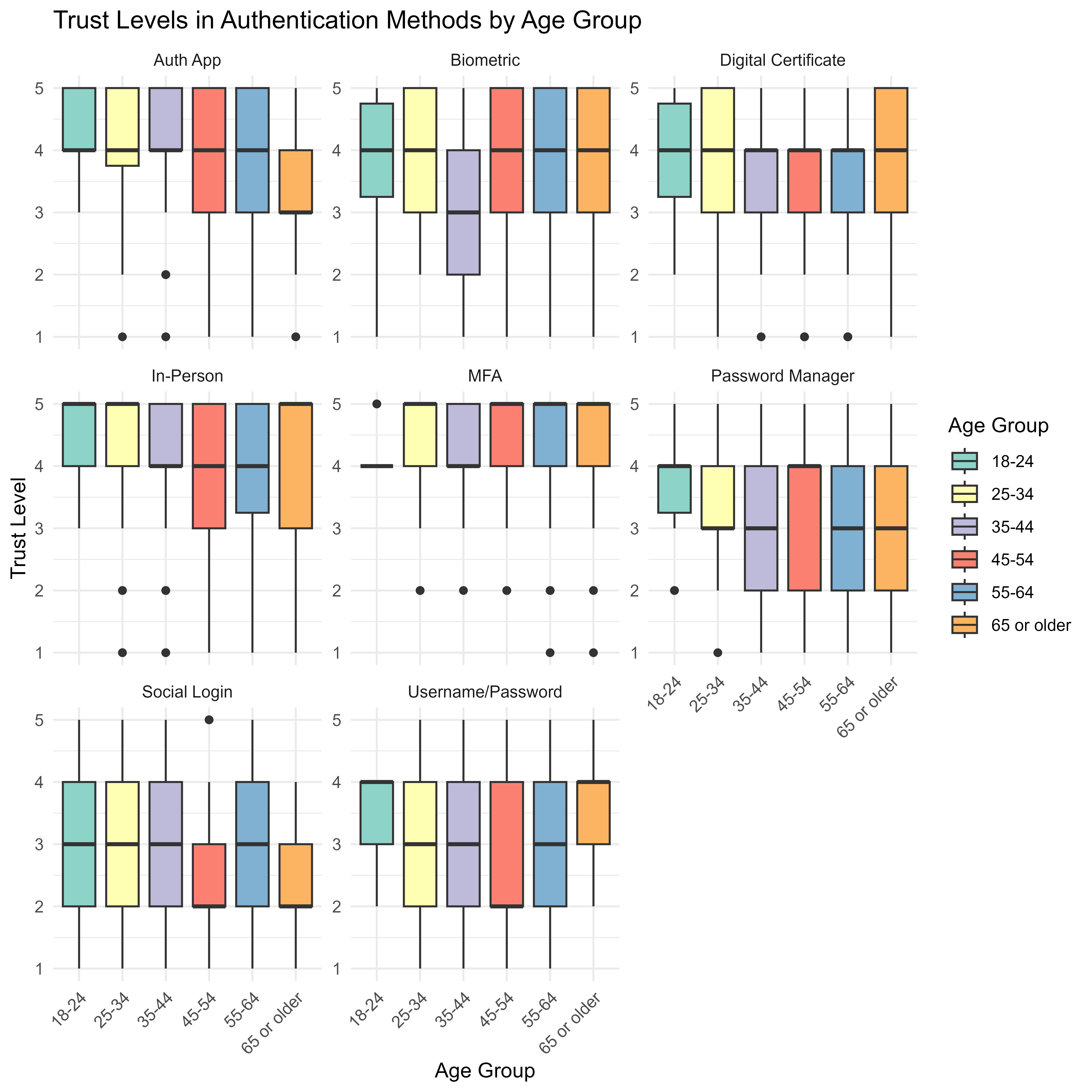}
    \caption{Average Trust By Age Group}
    \label{fig:trustByAge}
\end{figure}

\subsection{Trust by Employment Status}
\begin{figure}[ht]
    \centering
    \includegraphics[scale=0.4]{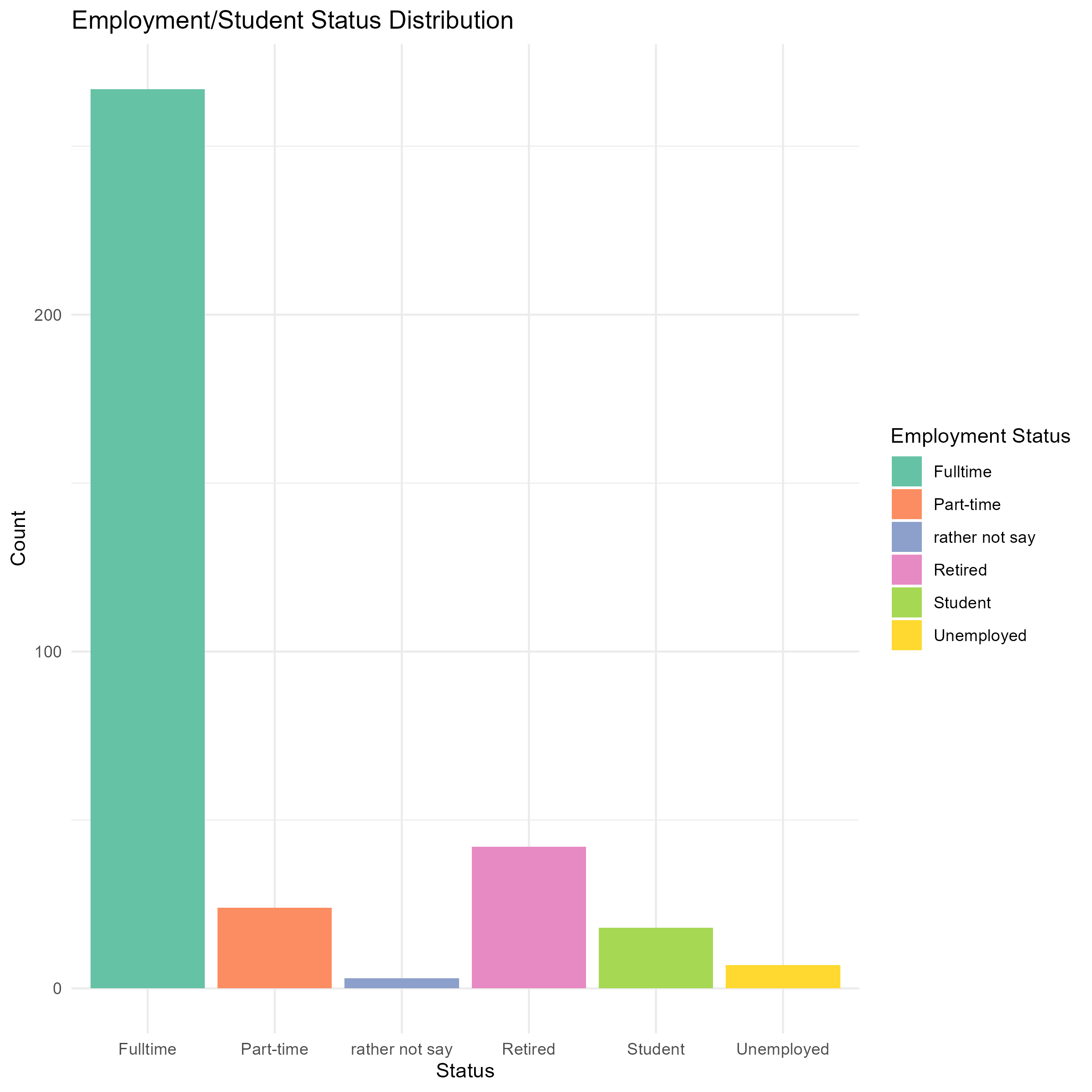}
    \caption{Employment Status Distribution}
    \label{fig:EmployDist}
\end{figure}

\begin{figure}[ht]
    \centering
    \includegraphics[scale=0.4]{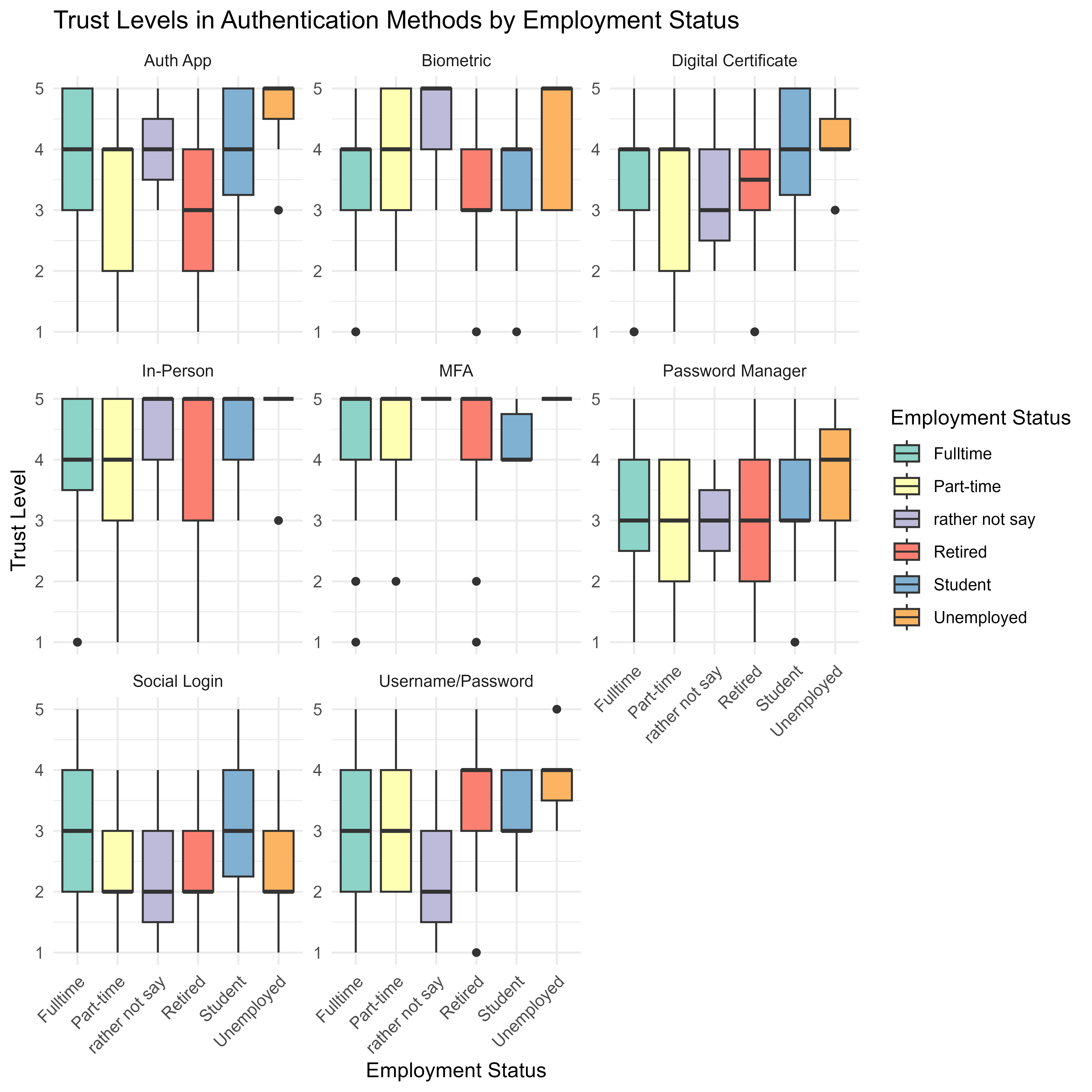}
    \caption{Average Trust By Employment Status}
    \label{fig:turstByEmp}

\end{figure}

Figure \ref{fig:EmployDist} shows the distribution of participants by their employment status. While the highest level of responses were received from full-time employees, which is representative of the employment status in Iceland, there was a small amount of retirees. This indicates that many of those aged 65 or older were in full-time employment. Moreover, the level of students responses was low, likely due to large numbers of students working along side their studies. It is important to take into consideration that the data collection was done during the summer, which potentially had an impact on the levels of responses from certain groups. 

Figure \ref{fig:TrustByGend} presents the trust by employment status in the authentication methods. Full-time and part-time employees generally exhibit higher and more consistent trust levels across most authentication methods. Students and unemployed individuals show lower and more varied trust levels, indicating less confidence in these authentication methods. Retired individuals show higher trust levels with minimal variability, similar to full-time and part-time employees.

\subsection{Trust by Gender}

Figure \ref{fig:genderDist} presents the distribution of participants by gender, illustrating more or less even representation of males and females. Figure \ref{fig:TrustByGend} presents the results of trust in authentication divided by gender. The data shows that female respondents have higher and more consistent trust levels across all authentication methods. In-person and \ac{MFA} methods are trusted consistently across female and male respondents. Social login, password manager and username/password methods have the lowest and most varied trust levels.

\begin{figure}[ht]
    \centering
    \includegraphics[scale=0.5]{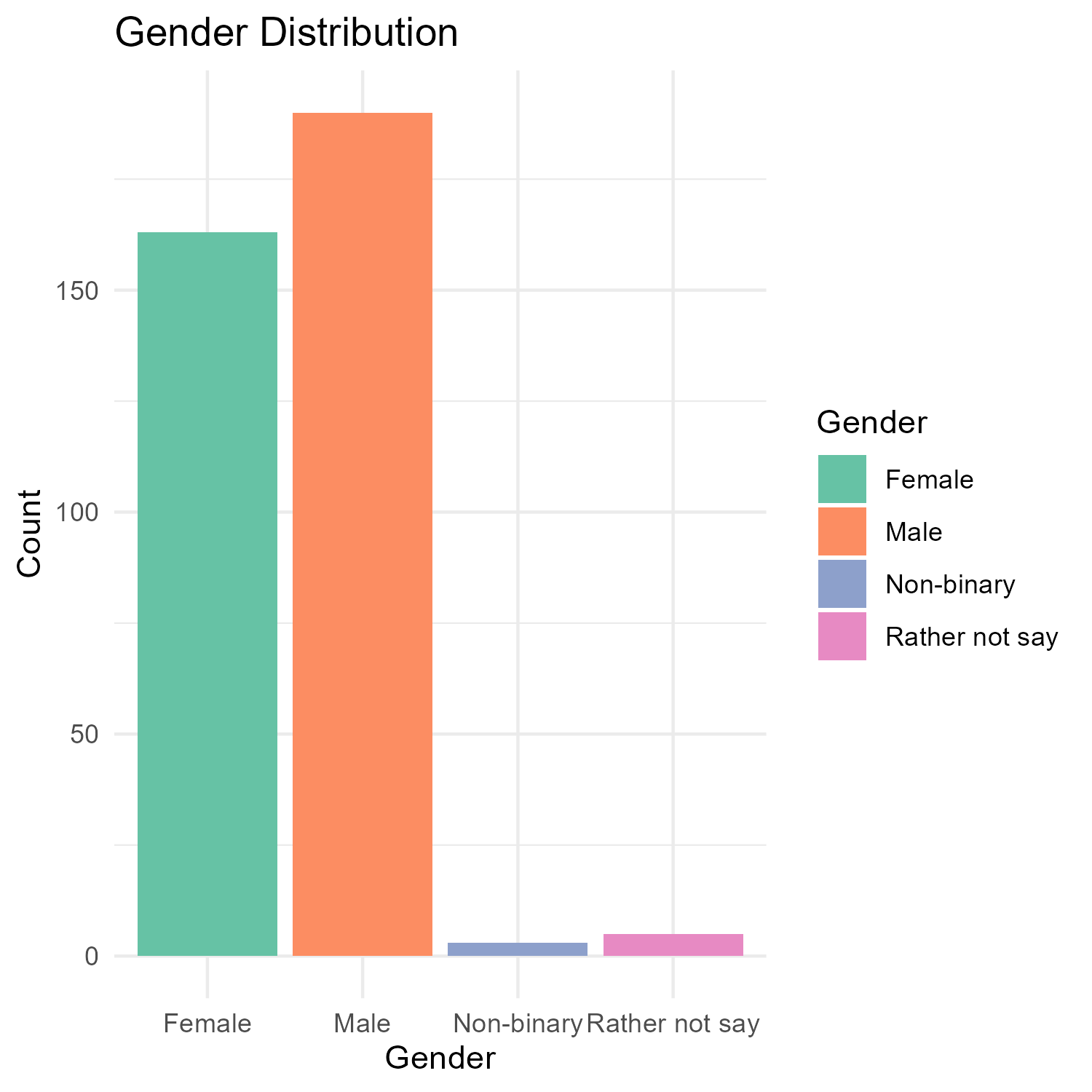}
    \caption{Gender Distribution}
    \label{fig:genderDist}
\end{figure}
\begin{figure}[ht]
    \centering
    \includegraphics[scale=0.4]{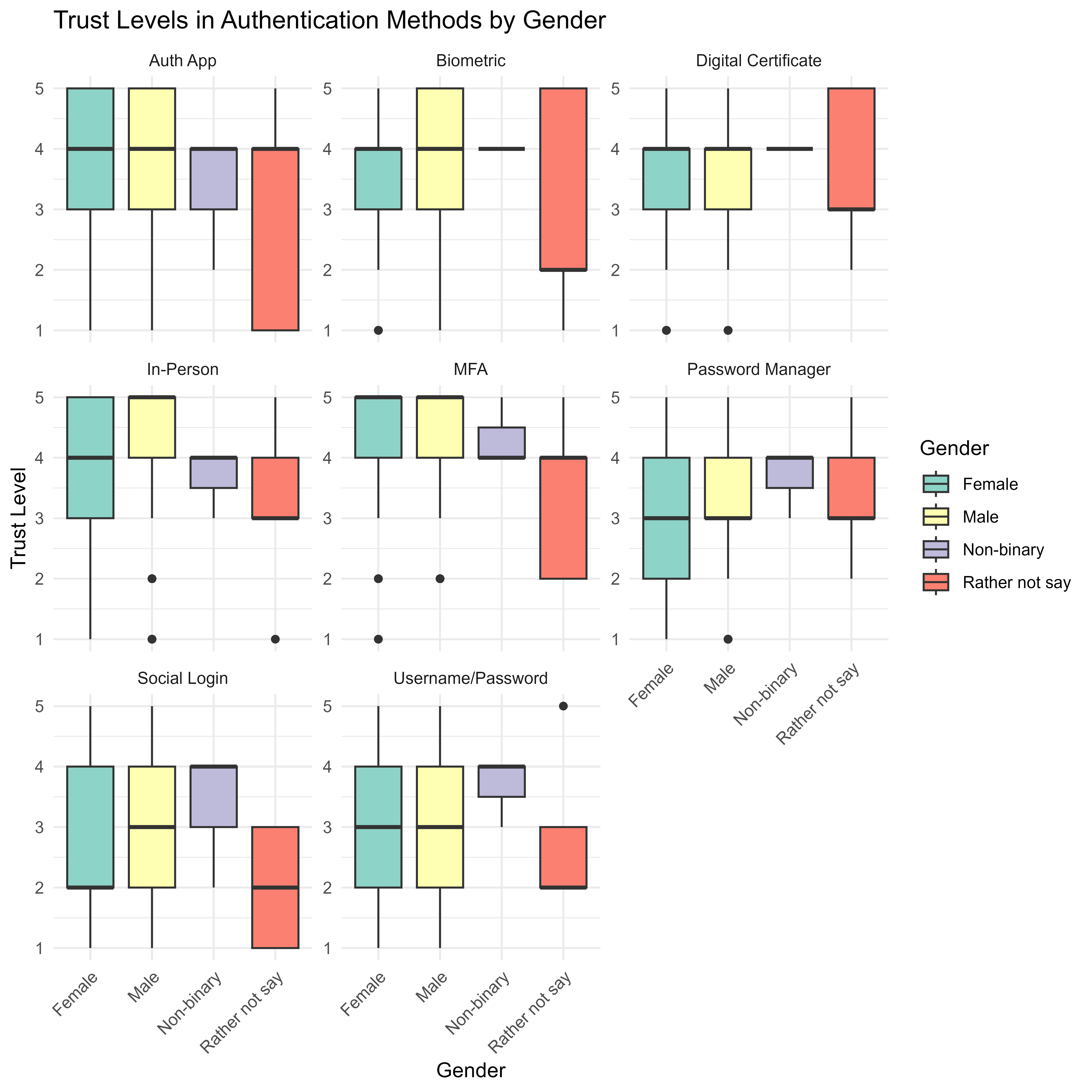}
    \caption{Average Trust By Gender}
    \label{fig:TrustByGend}
    \end{figure}

\subsection{Overall Trust}

\begin{figure}[ht]
    \centering
    \includegraphics[scale=0.45]{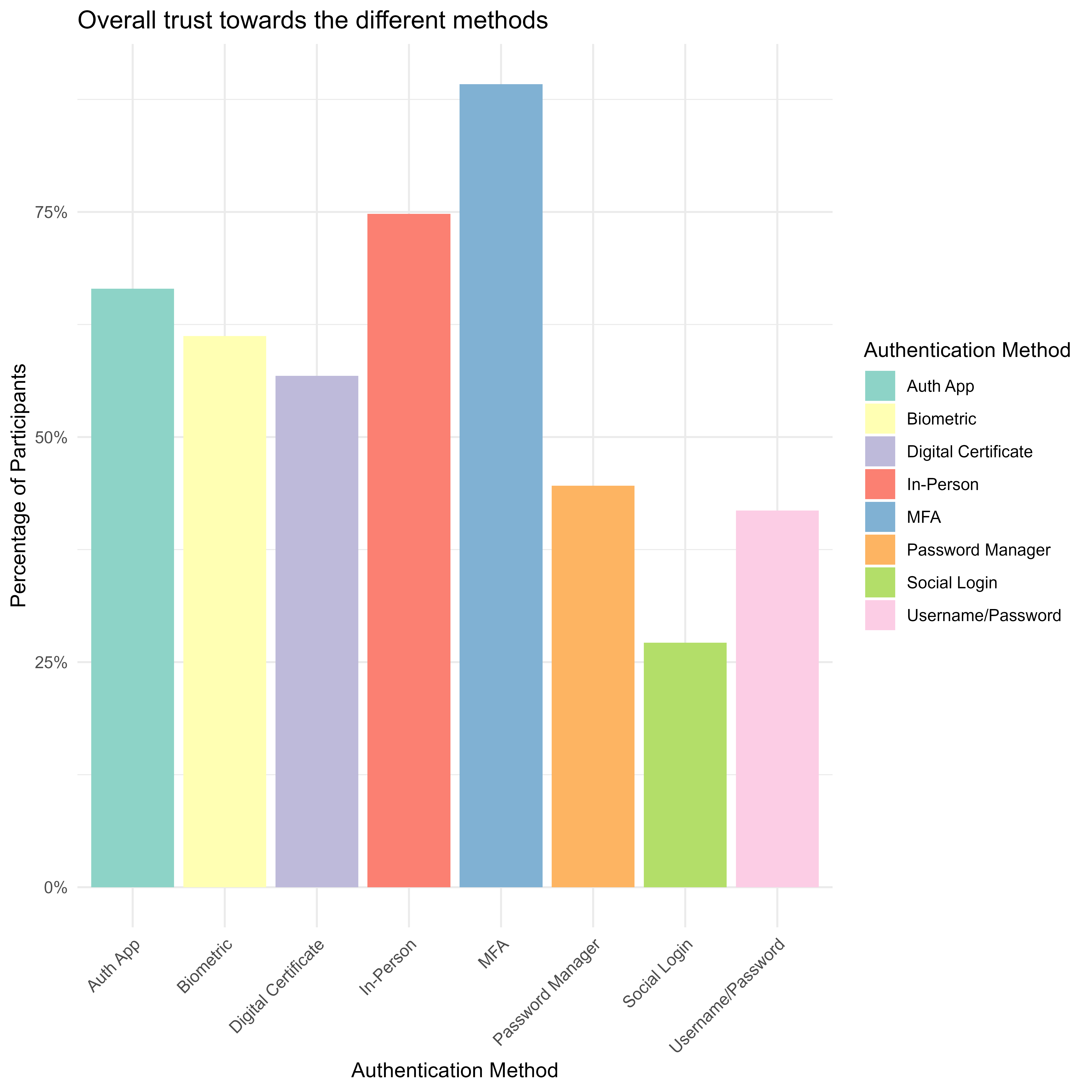}
    \caption{Average Overall Trust}
    \label{fig:overallTrust}

\end{figure}

This section presents the results across all demographics. Figure \ref{fig:overallTrust} illustrates the average trust in all authentication methods across demographics studied. The data shows that \ac{MFA} is the most trusted authentication method, followed by in-person visits. Respondents regard auth app, biometric and digital certificate as moderately trusted methods. However, respondents place lower levels of trust in password manager and username/password authentication methods while social login receives the least trust out of all considered methods.

\begin{figure}[ht]
    \centering
    \includegraphics[scale=0.4]{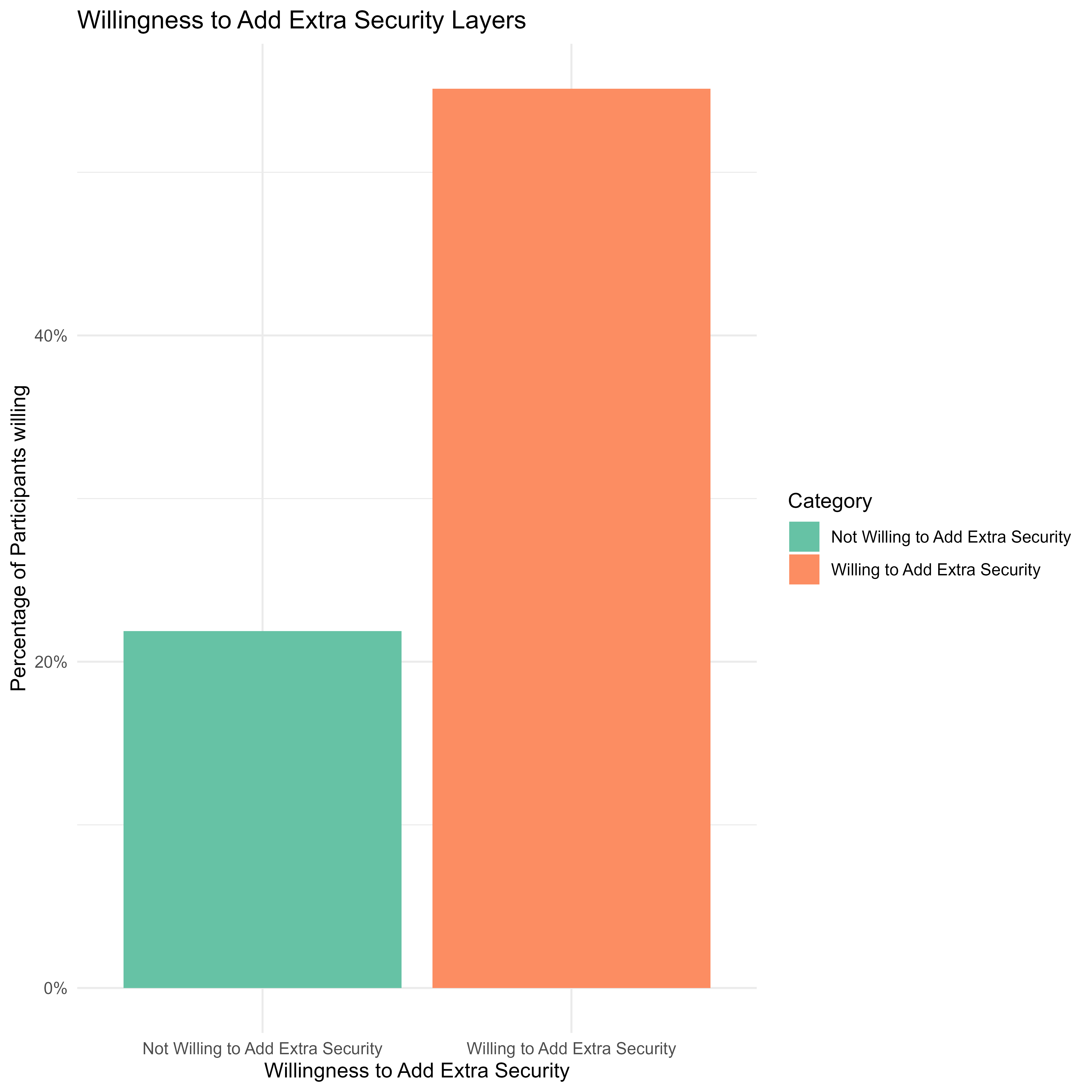}
    \caption{Overall Willingness for Additional Security steps}
    \label{fig:overallWill}
\end{figure}

Figure \ref{fig:overallWill} presents the answers to the question whether participants would be willing to increase the level of security in place even though it might cause more inconveniences. The data shows that across all demographics respondents demonstrate significant willingness to add further security layers to the authentication methods.

\section{Discussion}\label{sec:Discussion}

In terms of RQ1, the survey results indicate consistent levels of trust across demographics with younger age groups (\mbox{18-24} and \mbox{25-34} years of age) being more open to newer technologies such as bio-metrics or those which require sharing data with 3rd parties such as social logins. All but the older age groups have substantial trust in authentication apps which highlights a gap in usability or understanding for this group. In terms of employment status, there is more variability in trust across the different demographics and methods. As workplaces make use of a variety of digital collaboration and productivity tools this could highlight the gap in trust between individuals who more regularly use these methods versus those that are unemployed. Finally, the disparity between trust in authentication methods between female and male respondents may warrant further research, as it could reflect distinct security needs or concerns. Women, for instance, may demonstrate greater awareness of secure password practices due to heightened risks of stalking and domestic abuse \cite{maftei2023give, leukfeldt2019}. This could indicate a need for more targeted digital security education and awareness tailored to each group’s specific experiences and vulnerabilities.

In terms of RQ2, the 1st, 2nd and 3rd highest trusted authentication methods were \ac{MFA}, in-person, and authentication app respectfully. While both the \ac{MFA}, i.e.\ the Icelandic electronic ID, and authentication apps are technical solutions, it is remarkable that respondents trust them differently. This might be because the authentication app is developed by a 3rd party international organisation (e.g.\ Microsoft, Google), whereas the \ac{MFA} is an Icelandic solution used ubiquitously and promoted by the Government of Iceland. If we compare only the 1st and 2nd highest trusted authentication methods, i.e.\ \ac{MFA} using the Icelandic electronic ID and in-person, then we see a significant shift in public trust, where a relatively new digital authentication method is favoured over traditional face-to-face interactions. This phenomenon might initially appear counter intuitive, considering that physical presence has historically been the predominant mode of securing services and verifying identity. However, these results align with other studies suggesting that the increased reliance on and integration of online mediums have led to a corresponding rise in social distancing behaviours \cite{turkle2017alone, bessiere2008effects,kraut1998internet}. The growing comfort and familiarity with digital interactions likely contribute to the heightened trust in \ac{MFA}, reflecting broader societal trends towards digitisation and remote engagements. Similar as with the age and gender graphs, in-person and \ac{MFA} using the Icelandic electronic ID are the most trusted methods, while social login and username/passwords methods have the lowest and most varied trust levels.

In terms of RQ3, the survey data clearly shows that across all demographics the majority of respondents were willing to add additional security methods. This might indicate that, in general, the participants showed strong trust in digital authentication techniques indicating that future technologies may have strong uptake.

\section{Threats to Validity}\label{sec:Threats to Validity}

\noindent\textbf{Internal threats}  to validity centre around the high profile nature of the topic which might impact the opinion of the survey participants. The short time period for data collection meant that limited events could impact the results of trust in the different authentication methods as all authentication methods had been widely used for many years by the Icelandic general public. However, there is generally a heightened global cybersecurity situation which also impacts Iceland. During the data collection a cyberattack hit a major newspaper publisher in Iceland~\cite{mblrvakurMajor} which might have biased some participants' trust in digital authentication systems. However, high profile cyberattacks are on-going, such as the attack which took down a major Icelandic university for months which started in February 2024 \cite{mblHackersMight}. Therefore, although such events might influence perspectives on trust in digital authentication methods, we consider this to be a continuous process which should unlikely impact the results substantially. Additionally, the use of Microsoft Forms as a tool to survey the participants ensured consistency in data collection and its analysis.

\noindent\textbf{External threats} to validity in this study largely focus on the generalisability of the results. There is an inherent bias towards users with high digital skills as they were selected using online social media platforms. As of 2021 Iceland, however, tops the charts for internet usage in the \ac{EEA} and has the highest percentage of social media users in the \ac{EEA} at 98\% \cite{europaEurostatRegional}. This ensures that this digital bias is less relevant to Iceland, yet, these results may not be generalisable to other, less digitally active societies. However, the sample of the population could have been more diverse, e.g. there is under representation of the non-binary population as compared to other genders. Initially, the survey was sent out and accepted responses from the population of those under 18 years old. However, due to their low rate of participation, this population sample has been omitted from the analysis so as not to bias the results. Another bias might have been introduced by using the data collection tool, Microsoft Forms. Participants will have varying trust in such a large and widely known private organisation which would impact their willingness to provide answers. However, mitigating this is challenging as tools from similar providers would have a similar impact on trust. Additionally Microsoft Forms was suitable as it complied sufficiently with local data protection regulations.

\section{Future Research}\label{sec:futureresearch}
The results of this study have highlighted several interesting research areas which are described below:\\
\noindent\textbf{Longitudinal and Comparative studies.} While the work in this paper presents an initial study of trust in authentication services, further studies can be longitudinal. They might consider changing attitudes over time and in comparison with different nation states. This is particularly relevant given the rate of technological change e.g. the introduction of biometrics, societal change due to increasing digitalisation, and change in the geo-political system. However, such studies will also be hampered by this change, as the results may no longer be timely enough. Therefore correct selection of sampling times according to these specific influencing factors will be important.\\
\noindent\textbf{Usability Studies.} The results of this study indicated a generally high-level of trust in digital authentication systems for the population of Iceland. However it did not study the implementations'  usability, particularly for demographics who traditionally have less access to digital technologies, such as those with disabilities. These demographics may have been excluded from this survey due to this ineffective usability. Therefore further studies should be done to evaluate the extent of usability failings and highlight novel approaches for interaction design to ensure all members of society can make use of secure digital authentication. \\
\noindent\textbf{Biometrics.} The principal focus of this study was on software solutions for authentication. However, several participants suggested they would be willing to add additional authentication methods. One potential method is the use of biometrics which have benefits for replacing  ``something you have or know'' with ``something you are'' and thus mitigates many issues that users of authentication systems have. However biometric authentication systems are also with challenges, whereby depending upon the technology used, a subset of the population is unable to enrol due to physical constraints or changes in physical characteristics (e.g.\ the wearing down of fingerprints). Furthermore, hostile environments (e.g.\ dust particles) can reduce the effectiveness of biometrics and cause users to fall back to less secure methods. Additionally, there is an increase in privacy concerns around biometrics due to their storage of highly personal data. However, these all present interesting research opportunities into understanding the effectiveness of different technologies for whole populations/natural environments and also approaches to strengthening the privacy of their implementations on a nationwide scale.\\
\noindent\textbf{Policy Implications.} As digital authentication via electronic IDs becomes a common method for accessing sensitive platforms like online banking and healthcare systems, the potential for criminal exploitation has risen. The Icelandic Computer Emergency Response Team (CERT-IS) issued a warning in 2023 about organized spear-phishing attacks and social engineering tactics that target electronic IDs to access personal banking information \cite{certis2024}. From a policy advisory perspective, it is essential to assess whether public trust in certain authentication methods could inadvertently open paths for unauthorized access to critical systems. High-trust methods are particularly vulnerable to exploitation, as attackers may manipulate this trust through phishing and social engineering, often requiring only minimal information, such as a phone number, to gain access. These cases demonstrate the risks of relying on single-factor authentication for high-stakes information, emphasizing the need for robust \ac{MFA} protocols. It is thus valuable insight into further discussions on policy level that there is willingness to add additional security methods. Policy recommendations should consider mandating \ac{MFA} for accessing sensitive information systems to match security practices with the sensitivity of the data being protected.

\section{Summary and Outlook}\label{sec:Conclusion}
In summary, this paper presented the results of a study to understand the Icelandic general public's trust in various authentication methods for digital public services. The motivation for this work was in identifying if such methods were impacted by the low-level of institutional trust identified within the society. Overall, the results show a high-level of trust in advanced authentication methods, as government sponsored electronic ID based on \ac{MFA}, over in-person visits. With the majority of users even willing to increase the security. 

While tourist numbers are large in Iceland (in comparison to the number of inhabitants), tourists rarely get in contact with digital public services at the visiting country, therefore they were excluded from our survey. It can be hypothesised that their level of trust depends on their background from their home country and therefore their trust is as diverse as their origin is.

In future work, we aim to perform deeper studies into the usability of other authentication methods such as biometrics and those suitable for demographics with heightened accessibility needs correlated with the increasing diversity of Icelandic society.

\section*{Acknowledgement}

This project has received co-funding from the European Union's Digital Europe Programme under grant agreement no.\ 101127453 National Coordination Centre for Cybersecurity in Iceland and 101127307 Defend Iceland: Nationwide bug bounty platform and from The University of Iceland Research Fund.

\IEEEtriggeratref{20} 

\bibliographystyle{IEEEtran}
\bibliography{refs}

\end{document}